\documentclass[twocolumn,showpacs,showkeys,preprintnumbers,prl,amsmath,amssymb]{revtex4}
\usepackage[dvips]{graphicx}
\usepackage{dcolumn}
\usepackage{bm}
\usepackage{color} 

\newcommand{\pap}{$\parallel$ }

\newcommand{\tv}{$T_v$ }
\newcommand{\tvy}{$T_v$}

\newcommand{\be}{\begin{equation} }

\newcommand{\ene}{\end{equation}}

\newcommand{\eqr}{Eq.~\ref}

\newcommand{\figr}{Fig.~\ref}

\newcommand{\hcl}{$B_{c1}$ }

\newcommand{\hcuy}{$B_{c2}$}

\newcommand{\tc}{$T_{c}$ }

\newcommand{\tcy}{$T_{c}$}

\begin{document}
\preprint{Published in Phys. Rev. B {\bf 86}, 024521 (2012)}
\title{Thermally-activated dynamics of spontaneous perpendicular vortices\\
tuned by parallel magnetic fields in thin superconducting films}

\author{Milind N. Kunchur}
\email[Corresponding author email: ]{kunchur@sc.edu}
\homepage{http://www.physics.sc.edu/kunchur}
\author{Manlai Liang}
\affiliation{Department of Physics and Astronomy, University of South
Carolina, Columbia, SC 29208}
\author{Alex Gurevich}
\affiliation{Department of Physics, Old Dominion University,
Norfolk, VA 23529}

\begin{abstract}
We report magneto-transport measurements on a superconducting
molybdenum-germanium (MoGe) film of thickness $d$=50 nm in parallel magnetic
fields and show evidence of a transition from a Meissner state to
a resistive state of spontaneous 
perpendicular vortices generated by thermal fluctuations
above a certain temperature $T>T_v(B)$. Here $T_v$ appears to match the
vortex core explosion condition $d\approx 4.4\xi(T_v)$, where $\xi$ is the
coherence length.
For $T>T_v$, we observed that a nonlinear current-voltage ($IV$) 
response (Ohmic at low currents and the power law $V\propto I^\beta$ at 
higher $I$) is exponentially dependent on $B^2$.
We propose a model in which the resistive state at $T>T_v$ is due to 
thermally-activated hopping of spontaneous perpendicular vortices tuned by
the pairbreaking effect of the parallel $B$.
\end{abstract}


\maketitle
\section{Introduction}

The mixed state of Abrikosov vortices in type-II superconductors
exists between the lower and upper critical magnetic
fields, \hcl and \hcuy. In a film of thickness $d$ smaller than the
magnetic penetration depth $\lambda(T)$ but larger than
the coherence length $\xi(T)$, the lower critical field
$B^{\parallel}_{c1} = (2\phi_0/\pi d^2)\ln(d/\xi)$ parallel to the film surface
can well exceed the bulk \hcl \cite{aaa}. As $d$ decreases the vortex currents get squished by the film surfaces, so that
the vortex core becomes unstable and extends all the way across the film if the thickness becomes smaller
than $d = 4.4 \xi(T_v)$ \cite{likharev}. Such a ``core explosion''
transition of a vortex into a phase slip center can occur in thin films at
temperatures $T>T_v$ because the coherence length
$\xi(T)=\xi_0(1-T/T_c)^{-1/2}$ diverges at the critical temperature $T_c$. The
behavior of mesoscopic vortex structures in confined geometries has attracted much attention
both experimentally \cite{meso} and theoretically \cite{gl,vodolaz}.

Thermal fluctuations can radically change the behavior of vortices in thin films in a parallel field $B$.
For $B<B_{c1}$, parallel vortices are expelled, which would usually imply the Meissner state,
however, at $T>T_v$ a thin film can be in a resistive state if short
vortices perpendicular to the film surface are spontaneously generated by
thermal fluctuations, the energy of such vortices $\simeq d\phi_0^2/4\pi\mu_0\lambda^2$
being of the order of $k_BT$. Perpendicular vortices can appear either
through the Berezinskii-Kosterlitz-Thouless (BKT) unbinding of vortex-antivortex
pairs, or through the nucleation of single vortices at the
film edge (unbinding from their antivortex images
that then hop across the bridge) \cite{gv}. Here a parallel magnetic field can
be used to tune this transition, as will be shown below.

In this work we report transport measurements on amorphous MoGe
films which are a good model system for investigating
intrinsic flux dynamics because of their low bulk pinning and isotropic
nature \cite{unpinned,instability}. We observed a clear transition to a
thermally-activated resistive state due to hopping of spontaneous perpendicular
vortices, with the dynamics tuned by the pairbreaking effect of 
the {\it parallel} magnetic field. The latter is unusual
because parallel magnetic field does not interact with perpendicular 
vortices in the standard London theory. This transition was observed
at $T$ close to the core explosion transition temperature $T\approx T_v$,
above which
the resistance becomes strongly dependent on the parallel magnetic field.

\section{Experimental methods}

The electrical transport measurements were made both using continuous
dc signals (detected with standard digital voltmeters/nanovoltmeters) and
using pulsed signals (made
with an in-house built pulsed current source, preamplifier
circuitry, and a LeCroy model 9314A digital storage oscilloscope). The
pulse durations are on the order of 1 $\mu$s, and the pulse repetition
frequency is about 1 Hz, which reduces macroscopic heating of the film.

The cryostat was a Cryomech PT405 pulsed-tube closed-cycle
refrigerator that went down to about 3.2 K. 
It was fitted inside a
1.3 Tesla GMW 3475-50 water-cooled copper electromagnet mounted on a
calibrated turntable. 
Calibrated cernox and hall sensors monitored $T$ and $B$ respectively.
The accuracy of the in-film-plane alignment of the
magnetic field was $\theta = 0 \pm 0.025^o$. In the data section below we show
$R$ vs $\theta$ curves in the resistive state that arises from the hopping of 
spontaneous perpendicular vortices, which depends on the pairbreaking
effect of the parallel field, which in turn depends sensitively on the
alignment. This allows an accurate zero adjustment of the
in-film-plane angle. We will see below that $R$ has a $B^2$ dependence, 
consistent with a pairbreaking scenario; on the other hand if $B$ was
slightly tilted so as to produce field-induced perpendicular vortices,
then $R$ would be proportional to $B$ rather than $B^2$. Also the
observed $R$ has an Arrhenius temperature dependence consistent with the
hopping of spontaneous perpendicular vortices and not the motion of 
field induced perpendicular vortices. 

The MoGe microbridge of thickness $d=50$ nm, width $w=6 \mu$m, and
length $l=102 \mu$m was oriented so that the $B$ was parallel to the
film plane and perpendicular to current $I$.
The film was sputtered onto a silicon substrate with a 200 nm thick
oxide layer using an alloy target of atomic composition
Mo$_{0.79}$Ge$_{0.21}$. The deposition system had a base pressure of
$2 \times 10^{-7}$ Torr and the argon-gas working pressure was
maintained at 3 mTorr during the sputtering. The growth rate was
0.15 nm/s. The film was patterned using photolithography and argon
ion milling.

Our films had the following parameters which were measured independently:
\tcy=5.45 K, $R_n$=540 $\Omega$, $dB_{c2}/dT|_{T_c}$=-3.13 T/K,
and $d I^{2/3}_{d}/dT|_{T_c}$ = -0.0119 A$^{2/3}/$K.
Here $R_n$ is the normal-state resistance at $T_c$ and $I_d(T)$
is the depairing current near \tcy.
$B_{c2}(T)$, and hence $\xi$, were determined to high accuracy in
our earlier work \cite{unpinned} by fitting the entire resistive transition
$R(T,B)$ to the flux flow theory.
rather than simply looking at \tc shifts at some
resistive criterion such as $R=R_n/2$. The value of
 $I'^{2/3}_{d}$ and hence $\lambda$ were estimated by
taking the \tc shifts at $R=R_n/2$, as described in detail in Ref. \cite{pbreview}

In this work we measured the voltage-current characteristics of the films in the Ginzburg-Landau (GL)
region close to $T_c$ where $\lambda(t)=\lambda_0/\sqrt{1-t}$,
$\xi(t)=\xi_0/\sqrt{1-t}$, $B_{c2}(T)=B_{c20}(1-t)$,
$I_d(t)=I_{d0}(1-t)^{3/2}$, and $t=T/T_c$. Here
$\lambda_0=(\phi_0 dw/3\sqrt{3}\pi \mu_0 I_{d0}\xi_0)^{1/2}$ = 646 nm
and $\xi_0 = (\Phi_0/2\pi B_{c20})^{1/2} =$ 4.39 nm.
The large GL parameter $\kappa=\lambda_0/\xi_0=147$, indicates
a very dirty film.
The Pearl screening length $\Lambda(t)=2\lambda^2(t)/d =
16.7/(1-t)$ $\mu$m well exceeds $w$ for all $T$, so
the sheet current density $J=I/w$ is uniform over the bridge width.

Our previous reviews and other papers \cite{unpinned,instability,mplb,pbreview}
give further details about the measurement technique, thermal analysis,
and parameter determination.

\section{Experimental results and discussion}

\begin{figure}[ht]
\includegraphics[width=0.8\hsize]{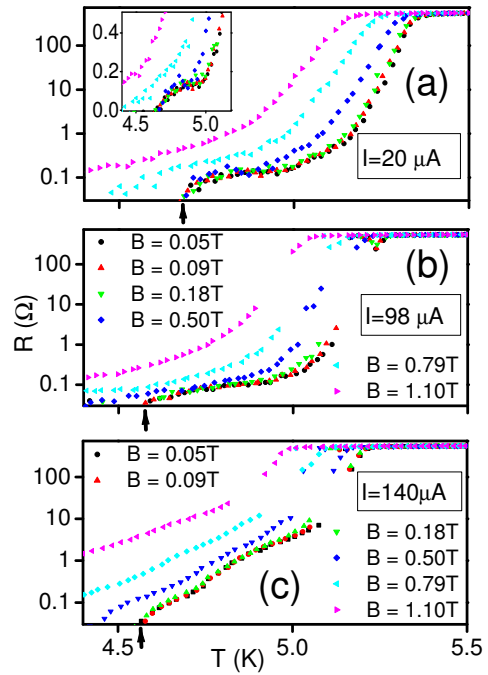}
\vspace{-1em} \caption{Resistive transitions in various indicated
parallel magnetic fields
and applied currents.
There is a transition temperature \tv
(indicated by arrows) at which
the $R(T)$ curves at low $B$ converge and plunge to zero.
$B$ values of different symbols are indicated in panels (b) and (c). The
inset of panel (a) shows its data plotted with linear axes for $R$ as
well as $T$.}
\label{RvsT-Bs}
\end{figure}

\begin{figure}[ht]
\includegraphics[width=0.8\hsize]{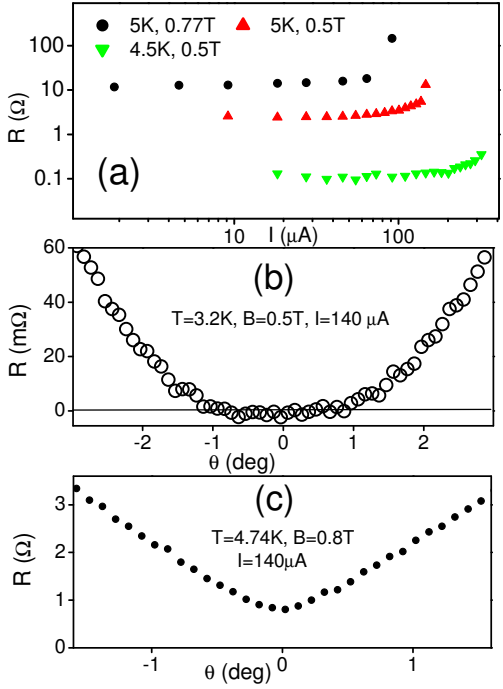}
\vspace{-1em} \caption{(a) Resistance versus current curves for a few
temperatures and fields above \tvy. The response is Ohmic at low $I$
turning non-linear at high $I$. (b) and (c) Angular dependence of
resistance at fixed $T$, $B$, and $I=140 \mu$A. (b) Below \tv there is an
$R=0$ plateau
for the angular range  $\sim \pm 2d/w \approx \pm 1^o$ within which the
applied \pap $B$ does not generate $\perp$ vortices. (c) Above \tv
$R\neq 0$ for all $I$ and $B$ because of the hopping of edge nucleated
$\perp$ vortices.}
\label{Ohmic}
\end{figure}

\figr{RvsT-Bs}
shows the temperature dependence of resistance observed at various magnetic
fields. Each panel corresponds to a different fixed current.
At lower $B$ and $I$, there is some kind of
transition temperature $T_v$ (marked by the arrows) at which the $R(T)$ curves
converge and plunge to zero.
Notice that the transition becomes less sharply defined
as $I$ and $B$ are increased. In the limit of low $I$ and $B$,
$T_v \simeq 4.7$ K.
There is a qualitative difference in the transport response above and
below \tvy. At $T>T_v$ a finite resistance was always observed:
an Ohmic response at low currents that becomes non-linear at high $I$
(\figr{Ohmic}(a)). Below \tvy, $R$ we observed zero resistance up to a high
value of $I$ 
on the order of $I_d$ at which an abrupt transition to the normal state occured.

The lower two panels of \figr{Ohmic} show
the dependence of $R$ on the angle between the applied $B$
and the film plane. For $T <T_v$, there is an $R=0$ plateau of angular
width corresponding to the tilt $\pm 2d/w \approx \pm 1^o$ that
causes the vortex to emerge outside the thickness; for  $T >T_v$, we
have $R\neq 0$ even at $\theta=0$ because of dissipation from
spontaneous perpendicular vortices generated by thermal fluctuations. 
As discussed in the `Experimental Methods' section, 
the angular dependence in this resistive regime allows the $\theta=0$ alignment
to be verified to high accuracy.

It turns out that the observed \tv is rather close to the core explosion
temperature $T_v$ defined by $d=4.4\xi(T_v)$ \cite{likharev}. Indeed, for
$\xi(T_v \simeq 4.7 \mbox{K}) = 11.8$ nm, we obtain 4.4$\xi=52$ nm, which is remarkably close to the film thickness $d = 50$ nm.
Phase transitions of the vortex matter (liquid-to-solid or
liquid-to-glass) can also cause sharp drops in $R(T)$ \cite{blatter}, but
in the field range of our measurements there can be no more than one row of vortices in our films.
Indeed, using the London expression for $B_{c1}^{\parallel}$ in which the vortex core energy is included \cite{gv}, we obtain
$B_{c1}^{\parallel}(4.7K) =(2\phi_0/\pi d^2)[\ln(d/\xi) - 0.07] = 0.72$ T, well
above the lowest magnetic fields at which the resistive transition at $T_v$ was observed. This estimate of $B_{c1}^{\parallel}$ may not be
very reliable for a film with $d\simeq 4\xi$, although more accurate calculations
of vortices in thin films in the GL theory have been done \cite{gl,vodolaz}.
Yet the London model suggests that the films ought to be in the Meissner state
at the lowest fields $B\simeq 0.05-0.1T$ of our measurements for which the
sharpest resistive transition at $T_v$ was observed.
Therefore, the motion of parallel vortices cannot explain the
resistance for $T>T_v$,
suggesting a transition to a resistive state in which short perpendicular vortices generated by thermal fluctuations hop across the film width.

The data above indicate that the film is in the Meissner state
for $T < T_v$ and that the resistive state above \tv 
may arise from thermally activated hopping of short perpendicular
vortices generated by thermal fluctations.
The activation barrier $U(T)$ is due to the variation of the self-energy of
the vortex across the film. In this case the $IV$ characteristic for
uncorrelated hopping of single vortices across the film of 
width $w<\Lambda$ and thickness $d\ll w$ was obtained Ref. \cite{gv}.
In the case of zero perpendicular magnetic field the $V-I$ characteristics
is given by
\cite{gv}
    \begin{equation}
    V=\frac{2IR_n(\beta-1)}{\gamma\Gamma(\beta+1)}\left[\frac{2\pi\xi}{w}\right]^\beta
    \bigl|\Gamma\bigl(1+\frac{\beta}{2}+i\gamma\bigr)\bigr|^2\!\sinh\pi\gamma,
    \label{exact}
    \end{equation}
where $\beta=\epsilon/T $, $\epsilon=d\phi_0^2/4\pi\mu_0\lambda^2$, $\gamma=\phi_0I/2\pi T$, and $\Gamma(x)$ is the gamma function.
Equation (\ref{exact}) gives an Ohmic $V(I) = R_vI$ at $I<I^*\xi/w$ and
a power-law $V(I)$ at $I^*\xi/w \ll I < I^*$, consistent with the observed
behavior of $V(I,T,V)$ shown in Figs. 1 and 2(a). Here $I^*=w\phi_0/2\pi
e\Lambda\xi$ is of the order of the depairing current. The asymptotic expressions for the resistance $R=V/I$ in these current domains are:
    \begin{gather}
    R(I)\simeq \sqrt{2\pi\beta}(I/I^*)^\beta R_n, \quad I^*\xi/w \ll I < I^*
    \label{uni} \\
    R_v\simeq \sqrt{2}R_n(\pi\beta)^{3/2}(\xi/w)^\beta, \quad I < I^*\xi/w.
    \label{ohm}
    \end{gather}
In the GL region $\beta(t)=\beta_0(1-t)/t$ with the Arrhenius
parameter $\beta_0=\phi_0^2d/4\pi \mu_0 \lambda_0^2k_BT_c \approx 436$, and
$I^*(t)=I^*_{0}(1-t)^{3/2}$ with $I^*_0=\phi_0dw/4\pi \mu_0 e\lambda_0^2\xi_0 \approx 7.9$ mA.

Before using Eq. (\ref{exact}) to describe our experimental data, we estimate
the range of temperatures $T_{BKT}<T<T_c$ of the BKT pair dissociation, where
$T_{BKT}$ is defined by the equation $\epsilon(T_{BKT})
=2k_BT_{BKT}$. Using the parameters
of our films presented above, we obtain that the BKT region $T_c-T_{BKT}= 2T_c/(2+\beta_0)\approx 0.025$K
is is very close to $T_c$ and is much narrower than the temperature range of our measurements.

\begin{figure}[ht]
\includegraphics[width=0.99\hsize]{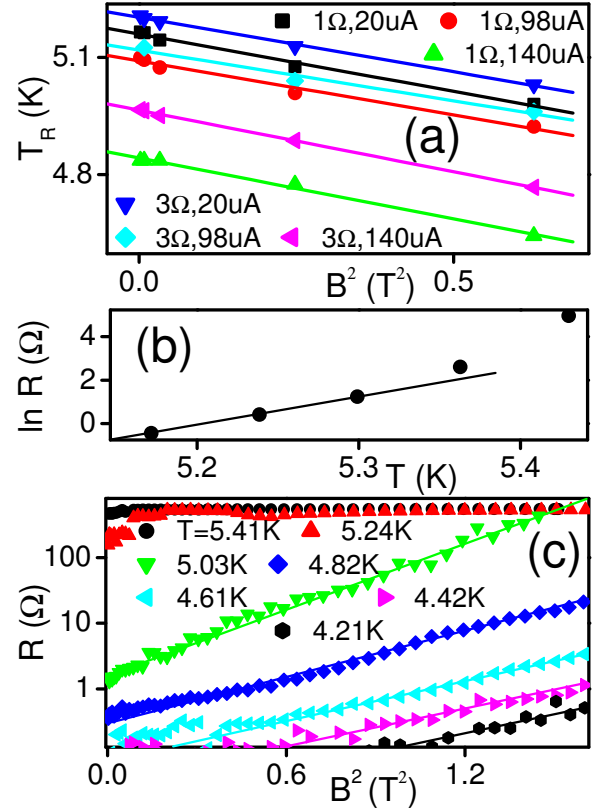}
\vspace{-1em} \caption{(a) Plots of $T_R$ versus $B^2$ for the indicated
resistance criteria and currents.
$T_R(R_c,B,I)$ is defined as the value of $T$ where $R=R_c$
on the curves in \figr{RvsT-Bs}, for the given $I$ and $B$. (b) 
Resistive transition at $I=13 \mu$A and $B$=0 used to obtain the zero-field
logarithmic slope $S$.
(c) The Ohmic resistance (at $I$=27 $\mu$A in middle of the Ohmic
current range as per \figr{Ohmic}(a)) shows an exponential dependence on
$B^2$. Different curves are at indicated temperatures.}
\label{RTvsB2}
\end{figure}

A parallel $B$ does not influence the dynamics of perpendicular vortices in
the London theory used to obtain
Eqs.~\ref{exact}--\ref{ohm}.
However, the GL pairbreaking of Meissner screening currents flowing parallel
to the film cause a variation in the large Arrhenius parameter
$\beta(T)=U/T$, which in turn leads to a strong $B$ dependence
of $R(T,B) \propto \exp (-U/T)$.
The Meissner currents reduce the
superfluid density $n_s(B)=n_s\int_{-d/2}^{d/2}(1-Q^2\xi^2)dx/d$
averaged over the film thickness, where $Q=2\pi Bx/\phi_0$ is the gauge-invariant
phase gradient of the order parameter \cite{nlme}. Thus, $n_s(B)=[1-(\pi\xi dB/\phi_0)^2/3]$ decreases
quadratically with $B$. Taking into account the gradient terms in the GL equation and assuming no suppression
of the order parameter at the film surface, we obtain a more accurate quadratic field correction in $\beta(B)$:
\begin{gather}
\beta(B)=\beta_0\left[1-(B/B^*)^2\right](1-t)/t,
\quad B\ll B^* \label{beta}  \\
B^*=\sqrt{3}\phi_0/\pi\xi\sqrt{12\xi^2+d^2}.
\label{h0}
\end{gather}
Here $B^*(T)$ is linear in $T$ at $T_c-T < 12T_c (\xi_0/d)^2$ and exhibits a square root
temperature dependence $B^*\sim \phi_0/d\xi \propto \sqrt{T_c-T}$ at
lower temperatures $T_c-T > 12T_c (\xi_0/d)^2\simeq 0.5$ K for our films.
Equations (\ref{uni}) and (\ref{beta}) predict an
exponential dependence of $R(B)$ on $B^2$:
\begin{equation}
t\frac{\ln(R_n/R)}{\beta_0\ln(I_0^*/I)}=1-t-\frac{B^2}{
B_0^2}\left[1+\frac{\alpha}{1-t}\right]
\label{int1}
\end{equation}
where $B_0=\sqrt{3}\phi_0/\pi\xi_0 d$, and $\alpha = 12(\xi_0/d)^2 \approx 0.093$. From Eq. (\ref{int1}) and in reference to the resistive transition curves
of \figr{RvsT-Bs}, we  define a ``critical temperature'' $T_R(R_c,B,I) =
t_R T_c$ at which $R(T,B,I)$ reaches a certain
value $R_c$ for given $B$ and $I$.
For $t < 1- \alpha$ (which holds for $R<10 \Omega$)
the last term in the brackets can be neglected,
and Eq. (\ref{int1}) gives $B = B_0 \sqrt{1 - t_Rf}$ where $f=1+\ln(R_n/R_c)/\{\beta_0\ln(I^*/I)\} \approx 1$.
Hence,
\begin{equation}
T_R \simeq T_c(1-B^2/B_0^2)/f
\label{hR}
\end{equation}
Eq. (\ref{hR}) shows that $T_R$ declines linearly with $B^2$ with a slope
$-T_c/fB^{*2}_0\approx -T_c/B^{*2}_0 =-0.22$. This estimate is
consistent with the average experimental
slope of -0.29 inferred from the plots of the measured $T_R$
(values of $T$ where
the curves in \figr{RvsT-Bs} attain 1 $\Omega$ and 3 $\Omega$ resistance
values) at the respective applied $B$ fields shown in \figr{RTvsB2}(a).
Eqs. (\ref{ohm}) and (\ref{beta}) give the exponential dependence of the Ohmic resistance $R_v$
on $B^2$:
\begin{gather}
\frac{d\ln R}{dB^{2}}=\frac{\beta _{0}}{tB_{0}^{2}}\left[ 1+\frac{\alpha }{%
1-t}\right] \ln \frac{w}{\xi (t)}. \label{slopeo}
\end{gather}

\figr{RTvsB2}(c) shows that the experimental Ohmic resistance
indeed varies exponentially with $B^2$ with a slope increasing with $T$.
(Note that if the perpendicular vortices were simply created by a
misalignment of the applied
field, then R would have been proportional to B instead of the $B^2$
dependence that comes from the mechanism we discussed.)
For the parameters of our films Eq. (\ref{slopeo}) gives
$d\ln R_v/dB^2$ about 10--12 times greater than the observed values,
mostly because of the factor
$\ln(w/\xi)\simeq 10$ in Eq. (\ref{slopeo}).
One reason for this discrepancy may be
the material uncertainty in $\beta _{0}$ resulting
from possible suppression of superconductivity at the film surface and
substrate.  The superfluid density can then
vary across the film even in the absence of the
parallel magnetic field. This effect is particularly pronounced at $T\approx T_{c}$
where $d<2\xi (T)$ so that any suppression of superconductivity at the
surface propagates all the way across the film.
To circumvent this
uncertainty, we use Eq. (\ref{ohm}) to express $\beta _{0}$ in terms of the
measured zero-field
slope $S=d\ln R/dt \simeq (\beta_0/t^2)\ln(w/\xi)$ at
$B=0$, shown in \figr{RTvsB2}(b).
Substituting $S$ into \eqr{slopeo} yields $d\ln R_v/dB^2$ in terms of measured parameters:
\begin{gather}
\frac{d\ln R}{dB^{2}}=\frac{St}{B_{0}^{2}}\left[ 1+\frac{\alpha
}{1-t}\right].
\label{lnR-B2}
\end{gather}
For $S=70.3$ inferred from the $\ln R$ vs $T$
data shown in \figr{RTvsB2}(b), \eqr{lnR-B2}
gives $d\ln R/dB^{2}=$ 2.83, 3.14, 3.52, 4.14, and 5.28 T$^{-2}$ at $T$
= 4.21, 4.42, 4.61, 4.82, and 5.03 K respectively. These are in agreement
within a factor of 1.5 with the measured slopes of \figr{RTvsB2}(c):
2.19, 2.21, 2.32, 2.69, and 3.85 T$^{-2}$.

These results indicate that the resistive state above \tv is consistent 
with thermally activated hopping of spontaneous perpendicular vortices 
tuned by the parallel magnetic field. Some of the quantitative discrepancy
with experiment may arise from randomly distributed 
pinning centers, 
which provide shorter hopping distances $\ell \ll w$.
Pinning centers in a film of thickness $d\sim\xi$
locally reduce the energy of a vortex by
$\alpha\epsilon$ where $\epsilon$ is the core energy and
$\alpha < 1$ depends on the details of pinning
interaction \cite{blatter}.
The drift velocity of the vortex $\bar{v}\sim \ell/\tau$
is then limited by the mean hopping time $\tau\propto\exp(\alpha\epsilon/T)$
leading to a much higher Ohmic resistance $R_v \sim
R_0\exp(-\alpha\epsilon/T)$ as compared to Eq. (\ref{ohm}) which implies
hopping of a vortex across the entire film width. Using
$\alpha\epsilon$
instead of $\epsilon\ln(w/\xi)$ in Eq. (\ref{slopeo}) significantly
reduces $d \ln R_v/dB^2$, in agreement with experiment. In turn,
pinning becomes less essential for the nonlinear part of
the $I-V$ curve where the activation barrier $U=\epsilon\ln(I_*/I)$ is not
only much reduced but also localized in a narrow $(\ll w)$ region at the
film edge \cite{gv}. As a result, Eqs. (\ref{int1}) and (\ref{hR}) describe
the experimental data well at high currents.

\begin{figure}[ht]
\includegraphics[width=0.8\hsize]{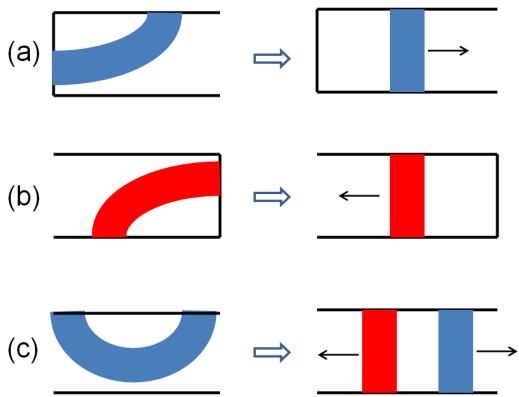}
\vspace{-1em} \caption{A sketch of vortex segments nucleated 
by thermal fluctuations at the edge of the film
(a) and (b) and on the broad face of the film (c).}
\label{edge}
\end{figure}
A possible mechanism by which the core explosion of parallel vortices could facilitate nucleation
of perpendicular vortices is illustrated by Fig. 4. There are three different possibilities: vortex quarter-loops of different polarities nucleate at the film edges (a and b) or vortex semi-loops nucleate at the broad face of the film (c). Because the normal component of circulating vortex current vanishes at the film surface, the energy $U_{ql}$ of the vortex quarter-loops at the edges is exactly half of the energy of the face semi-loop $U_{sl}=2U_{ql}$ which can be regarded as two quarter-loops with opposite polarity. A low temperatures the edge quarter-loops will therefore dominate the thermally-activated voltage $V\propto\exp[-U_{ql}/k_BT]$ despite smaller statistical weight of edge nucleation as compared to the semi-loop nucleation on the broad face of the film. As temperature increases, particularly at $T$ close to $T_{BKT}$, the semi-loop nucleation takes over and $V$ is determined by the BKT pair dissociation.
In thin films with $d<4.4\xi$, the segments of vortex semi-loops or quarter-loops parallel to the broad face of the film
are unstable and quickly propagate across the film. As a result, the edge semi-loops turn into either short perpendicular vortices (a) or antivortices (b) which are then driven by transport current across the film. The expansion of vortex semi-loop in Fig. 4c turns it into vortex-antivortex pair which are then driven apart by current. This process is a part of the BKT pair dissociation. Detailed calculation of these processes requires numerical simulation of the 3D Ginzburg-Landau equations (see, e.g. Ref. \cite{vodolaz}).

\section{Conclusion}

To summarize, our magneto-transport measurements in a thin MoGe film in a
parallel magnetic field revealed a thermal fluctuation-driven transition to a
dissipative state caused by motion of thermally-activated perpendicular
vortices. This results in the Arrhenius $V-I$ characteristics which have been
observed in several decades in voltage. It appears that
the observed transition temperature\tv matches the vortex explosion condition
of $d=4.4\xi(T)$.
We showed that the resistive state above \tv can be tuned by the
pairbreaking effect of the parallel field. The suppression of the sheet 
superfluid density by the parallel field can also be used to tune the $V-I$
characteristics of thin superconducting films in the BKT
temperature region.

The authors gratefully acknowledge Jiong Hua, Zhili Xiao, James M.
Knight, and Richard A. Webb. This work was supported by the U. S. Department
of Energy through grant number DE-FG02-99ER45763.

\end{document}